# Dosimetric Evaluation of MapCHECK in MapPHAN for Conformal Arc SABR Quality Assurance

**Dr Nathan I. N. Henry – corresponding author**

Affiliations:

1. University of Western Australia, Perth, Australia

ORCID: 0000-0002-4299-1442

Email: nathan.henry@uwa.edu.au

**Christopher M. Thompson**

Affiliations:

1. Department of Medical Physics, Leeds Cancer Centre, Leeds Teaching Hospitals NHS Trust, Leeds, UK

ORCID: 0009-0009-1389-6663

**Dr Steven Marsh**

Affiliations:

1. University of Canterbury, New Zealand

ORCID: 0000-0001-7007-4759

**Dr Jack D. Aylward**

Affiliations:

1. Medical Physics, Bristol Cancer Institute, Bristol, UK
2. Medical Physics, School of Applied Sciences, University of the West of England, Bristol, UK

ORCID: 0000-0002-0801-228X

# Abstract

This study investigates the sensitivity and specificity of the MapCHECK phantom housed in MapPHAN to errors in conformal arc Stereotactic Ablative Body Radiotherapy (SABR). A specific focus is applied to Organ At Risk (OAR) dosimetry. Multi-Leaf Collimator (MLC) class shift errors up to 2 mm, and isocentre shift errors of up to 1 mm, were introduced to 23 simulated 6 MV X-ray conformal arc lung SABR plans in Raystation. Overall, 198 plans were delivered on a Varian Clinac iX linear accelerator to an in-house modified MapCHECK in MapPHAN. Gamma analysis was used to compare the MapCHECK measurements to the simulated Raystation plans. Receiver Operating Characteristic (ROC) curves were generated from these results to determine the sensitivity and specificity of the measurement technique to the introduced errors. For MapCHECK in MapPHAN, a combination of 5%/1 mm with 95% gamma tolerance and 2%/1 mm with 90% tolerance provides good sensitivity and specificity for Quality Assurance (QA) of conformal arc SABR plans.

# Introduction

Much work, mainly focussing on Dose Volume Histogram (DVH) metrics related to Planning Target Volume (PTV) doses, has been done to analyse the sensitivity and specificity of Quality Assurance (QA) phantoms to introduced errors [1]–[8]. For example, Kim *et al.* [1] performed a Receiver Operating Characteristic (ROC) analysis of gamma index sensitivity to introduced multileaf collimator (MLC) errors, but did not focus on Organ At Risk (OAR) doses. Further, Fredh *et al.* [4] evaluated the performance of multiple QA systems to introduced errors, comparing gamma analysis results and DVH for both target and OAR doses on a number of sites, for plans with and without introduced errors. However, little has been done to determine the sensitivity and specificity of QA methods to introduced errors that affect OAR doses, in particular for Stereotactic Ablative Body Radiotherapy (SABR) treatments.

SABR, also known as Stereotactic Body Radiotherapy (SBRT), utilises high dose-gradient techniques to deliver high dose-per-fraction treatment with sub-millimetre precision [9]. In order to validate the accuracy of these treatments, a QA solution with equivalent resolution and accuracy is required. Film provides an adequate, if time-consuming, solution for this task [10]. However, as SABR treatments become more prevalent, there is a growing demand for a more efficient solution. Diode arrays are an alternative option, with lower resolution and an angular dependence to radiation dose that must be accounted for during the commissioning process. However, they provide instantaneous dose readout, which is a significant advantage over film [11].

The diode array used in this work is that within the MapCHECK 2 phantom, model 1177 (Sun Nuclear Corporation, Melbourne, Florida), consisting of 1527 N-type SunPoint diode detectors, uniformly spaced 1 cm apart within a 32.0 x 26.0 cm$^2$ 2D array, and housed within an acrylic casing. Sun Nuclear provides an optional solid water case known as the MapPHAN, which allows MapCHECK to be used for QA of arc therapy plans. It has been shown that the angular dependency of the diodes in this phantom can be reduced to within ± 2% by applying density corrections to the phantom model *in silico* [11]–[13].

This paper reports on analysis to determine whether the MapCHECK in MapPHAN can be used to identify common treatment errors with sufficient accuracy to replace film as a regular QA technique.

In particular, the ability of MapCHECK in MapPHAN to accurately predict which conformal arc SABR plans will exceed OAR dose tolerances was analysed. To the authors' knowledge, this has not been previously reported.

# Methods & Materials

### In silico analysis of introduced errors

Twenty-three lung cancer patients treated with conformal arc lung SABR at Auckland Hospital were selected for inclusion in this study. Of these, fourteen patients received a minimum of 48 Gy to the PTV in four fractions, and nine had a minimum of 60 Gy delivered to the PTV in eight fractions. All plans were simulated using 6 MV flattened X-ray beams on a Clinac iX (Varian Medical Systems Inc., CA, U.S.A.) equipped with a Millenium 120 MLC. Each plan was optimised in Pinnacle v9.8 or v9.10 (Koninklijke Philips N.V., Amsterdam, Netherlands). Patients were scanned with a Siemens Somatom Sensation CT with hands raised above their heads in the supine position. The Varian Real-Time Position Management Gating System (Varian Medical Systems Inc., CA, U.S.A.) was used to generate an ITV via reconstruction of eight breathing phases captured in CT. An isometric margin of 5 mm was then added to the ITV volume to generate the PTV contour. The arc length for each plan was in the range of 180º-220º, with each plan having either two or three arcs, and 5º spacing between control points. The plans for these patients were anonymised in Pinnacle and imported into Raystation v5.0 (Raysearch Laboratories, Stockholm, Sweden) to manually edit plan parameters and compare calculated plans. Control point spacing was manually edited to 4º in Pinnacle to facilitate plan transfer into Raystation.

Systematic class open MLC shifts were introduced to the treatment plans. Python 2.7 [14] was used to introduce incremental class open errors of 0.2 mm (up to a maximum of 1 mm), widening the MLC aperture for all leaves on both MLC banks. A dose grid voxel size of 0.2 x 0.2 x 0.2 cm$^3$ was used to calculate both the original and modified plans. In a similar fashion, Raystation was used to introduce 1 mm isocentre shift errors, with the isocentre position shifted both ways along all three cardinal axes - inferior-superior, right-left, and anterior-posterior.

Following these edits, the maximum point dose for each contoured organ within the plan was calculated in Raystation and recorded. These values were plotted as a function of the magnitude of introduced MLC and isocentre shift errors. The gradients of these plots, determined via linear regression, were then used to determine the significance of the introduced errors.

### Commissioning of MapPHAN

The MapCHECK 2, Model 1177 phantom was mounted between two slabs of solid water which comprised part of a MapPHAN-MC1 accessory, previously designed for use with the MapCHECK 1 phantom (Model 1175). To fit the MapPHAN accessory to the MapCHECK 2 phantom, two solid water inserts with dimensions of 11 mm x 29 mm x 300 mm were created in-house and fitted between the two MapPHAN slabs. This was done so the inserts abutted the left and right sides of the MapPHAN phantom, as seen in Figure 1. The phantom was scanned using a Siemens Somatom Sensation CT scanner, in both coronal and sagittal orientations, and the scans were imported into Raystation.

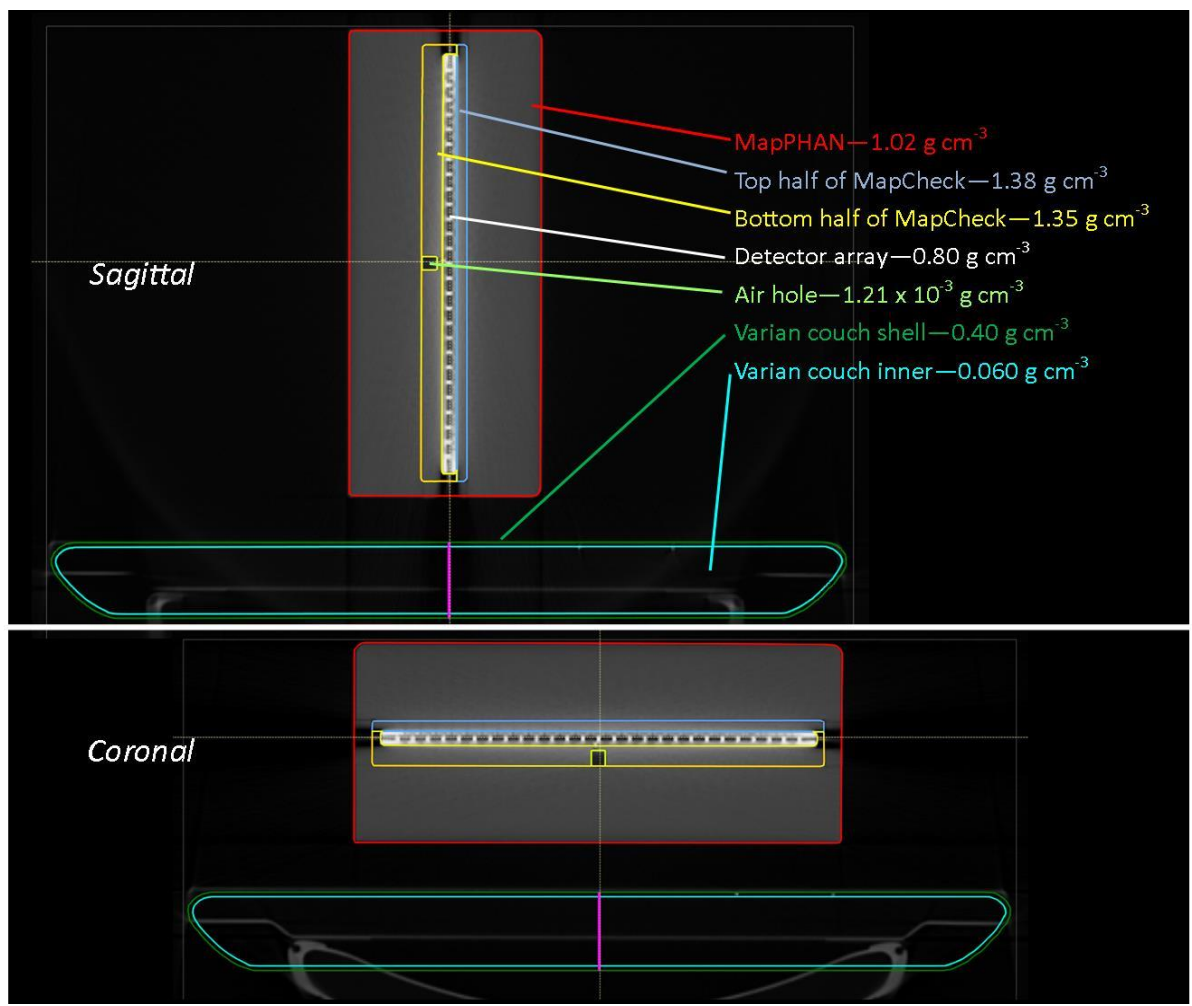

*Fig. 1* CT scans of MapCHECK in MapPHAN, in both coronal and sagittal orientations, with density overrides applied for reduction of angular dependencies. Streak artefacts can be observed in the plane of the detector array

Density overrides were applied to all sections of the phantom in coronal orientation to reduce the angular dependency to acceptable tolerances, due to image artefacts in the plane of the detector array. These overrides were optimised to minimise the disparity between angular dependency measurements and Raystation calculated doses. The overrides, shown in Figure 1, were then applied without modification to the phantom model in both orientations.

## *Measurements*

The treatment plans from 9 patients (both original and with MLC errors applied) were delivered to the phantom using a Varian Clinac iX – in total, 198 plans were generated for the final analysis. The modified DICOM files for each patient plan were exported directly from Raystation. These plans were used to calculate dose distributions on the simulated MapCHECK in MapPHAN phantom in Raystation, using both the original plans, and plans with MLC shifts included. The RTPlan and RTDose DICOM files for these plans were analysed in Sun Nuclear (SNC) Patient software.

Six gamma analysis settings were used: 1%/1 mm, 2%/1 mm, 2%/2 mm, 3%/2 mm, 3%/3 mm, and 5%/1 mm. A gamma threshold of 10% was set for all measurements. Two measurements were taken for each plan, with the phantom offset by 5 mm (half the detector spacing) superior via couch shift for the second measurement. These two measurements were then merged in SNC Patient, thus doubling the phantom's effective detector resolution. Following this, shifts were applied to align the measured and calculated dose distributions. For the isocentre shift results, gamma analysis was performed both with and without automatic shift corrections included.

In order to determine the sensitivity and specificity of the MapCHECK phantom to the introduced plan errors, ROC curves were generated using the 'ROCR' package [15] in the R programming language [16]. The Area Under Curve (AUC) and Youden index were used to quantify the sensitivity and specificity of the phantom for each set of gamma indices.

# Results

Using the density overrides shown in Figure 1, the mean angular dependency was reduced to within ± 2% for the central diodes of the phantom, with the highest discrepancy being observed when the diodes were irradiated laterally. This effect can be attributed to photon starvation effects that were observed during CT, where streak artefacts were seen in the array plane, as can be seen in Figure 1.

### *In silico analysis*

Figures 2-5 contain the gradient data generated from the introduction of MLC and isocentre shift errors. In these plots, the initial maximum point dose was measured from the original plan calculated in Raystation, while the gradient was measured from linear trendlines that were fitted to the data as a function of the introduced errors. The slanting dashed lines indicate the OAR dose tolerances for these plans, adjusted for the maximum point dose gradients as a function of the introduced errors.

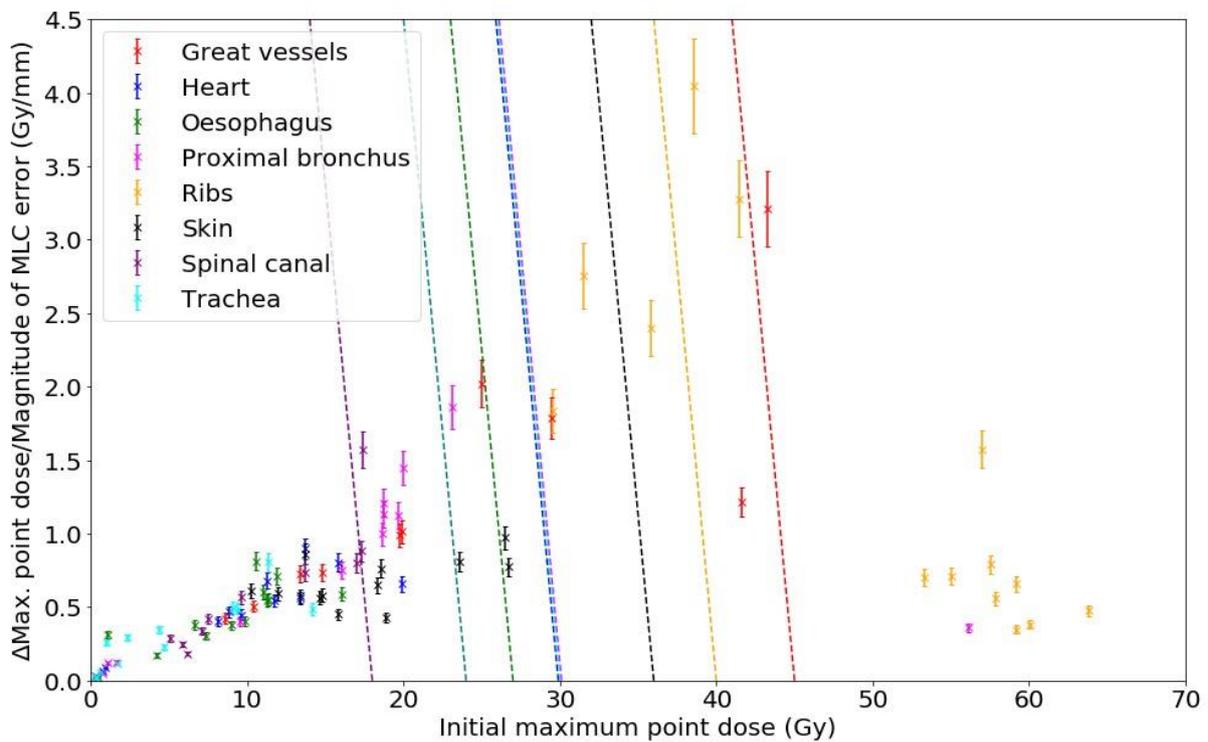

**Fig. 2:** *Point dose gradients induced by MLC class-open shift errors, with PTV receiving at least 48 Gy*

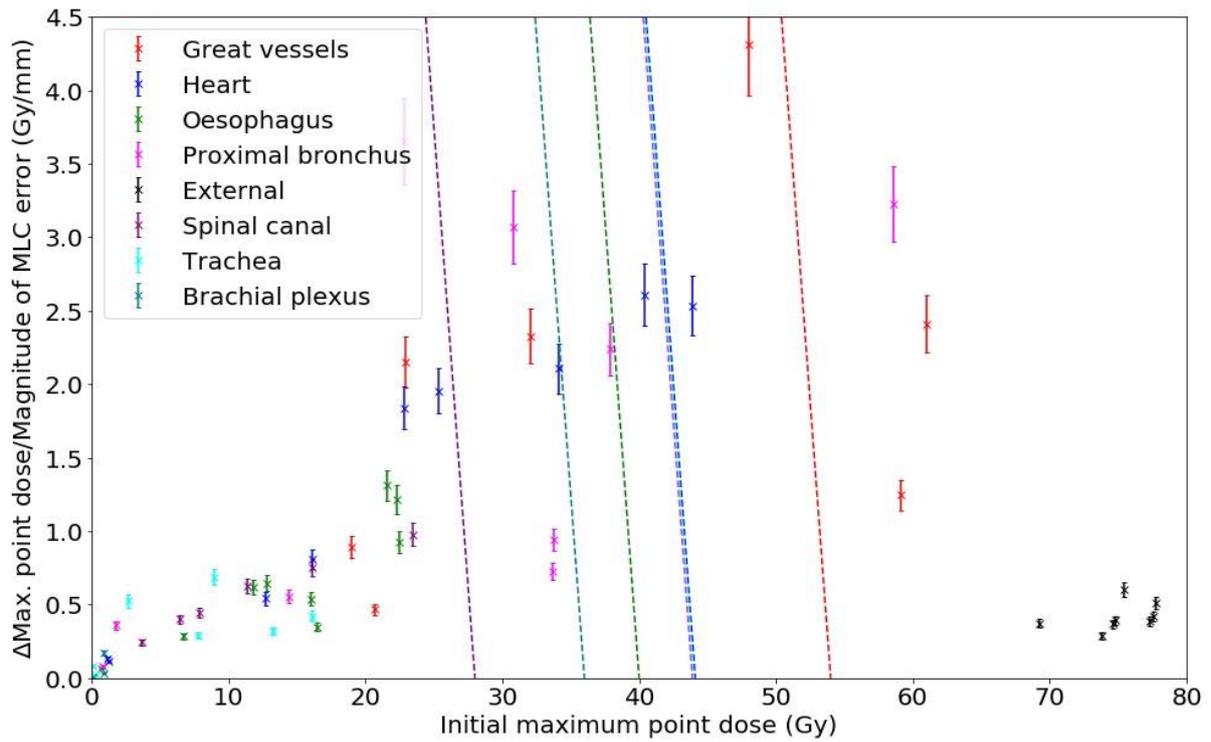

**Fig. 3** *Point dose gradients induced by MLC class-open shift errors, with PTV receiving at least 60 Gy*

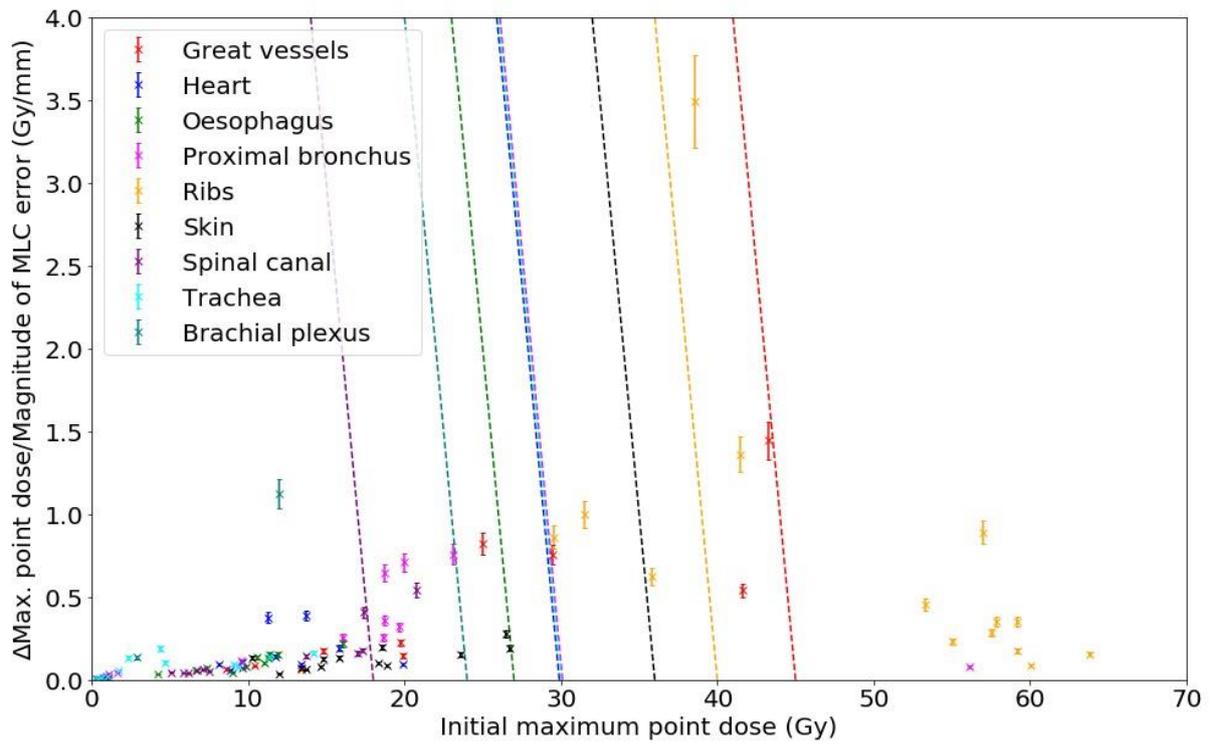

**Fig. 4** *Point dose gradients induced by isocentre shift errors, with PTV receiving at least 48 Gy*

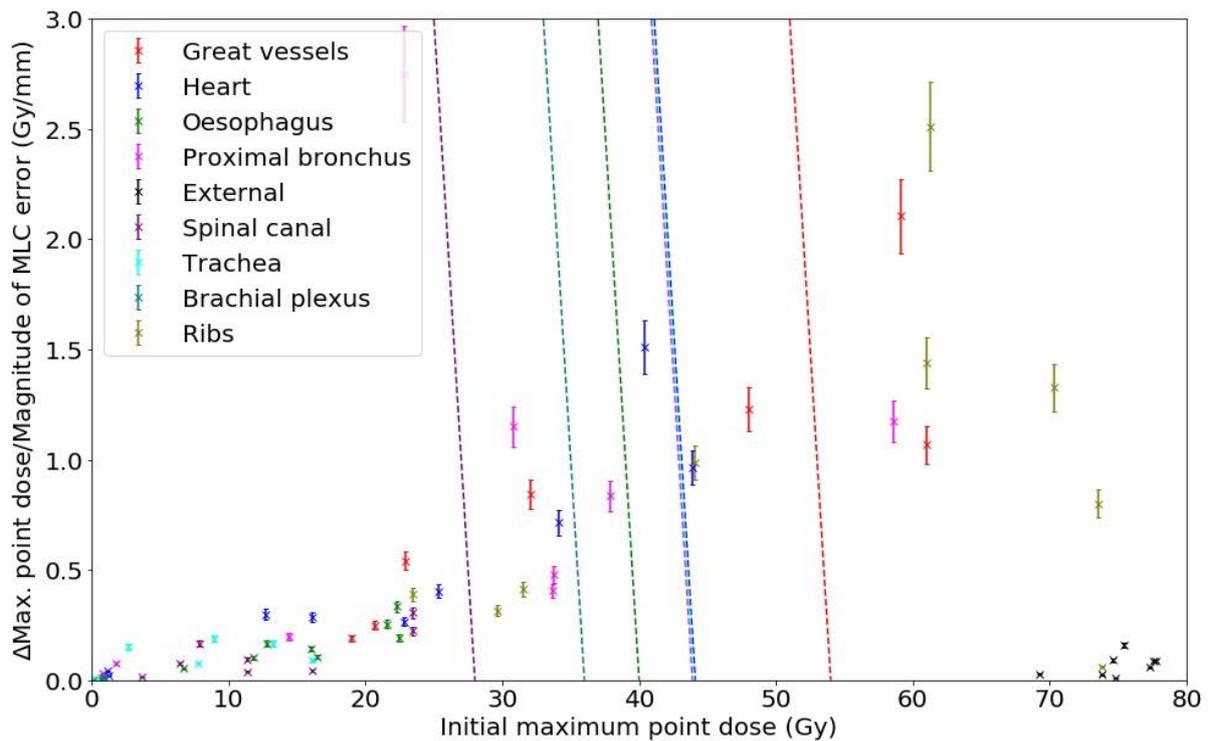

**Fig. 5** *Point dose gradients induced by isocentre shift errors, with PTV receiving at least 60 Gy*

### PSQA measurements

The mean gamma pass rates for each type of introduced error are presented in Figures 6-7. For each gamma tolerance, the inclusion of shifts in SNC Patient increased the mean pass rate for the isocentre shift results, and the mean coronal gamma pass rate was lower than the corresponding mean sagittal pass rate.

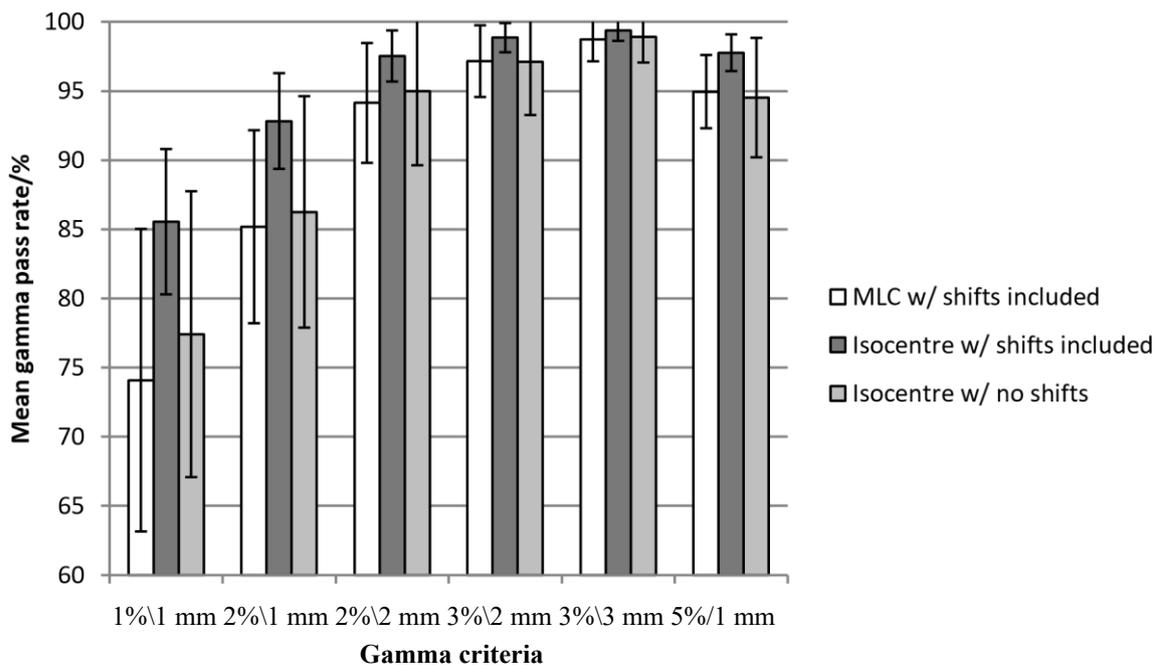

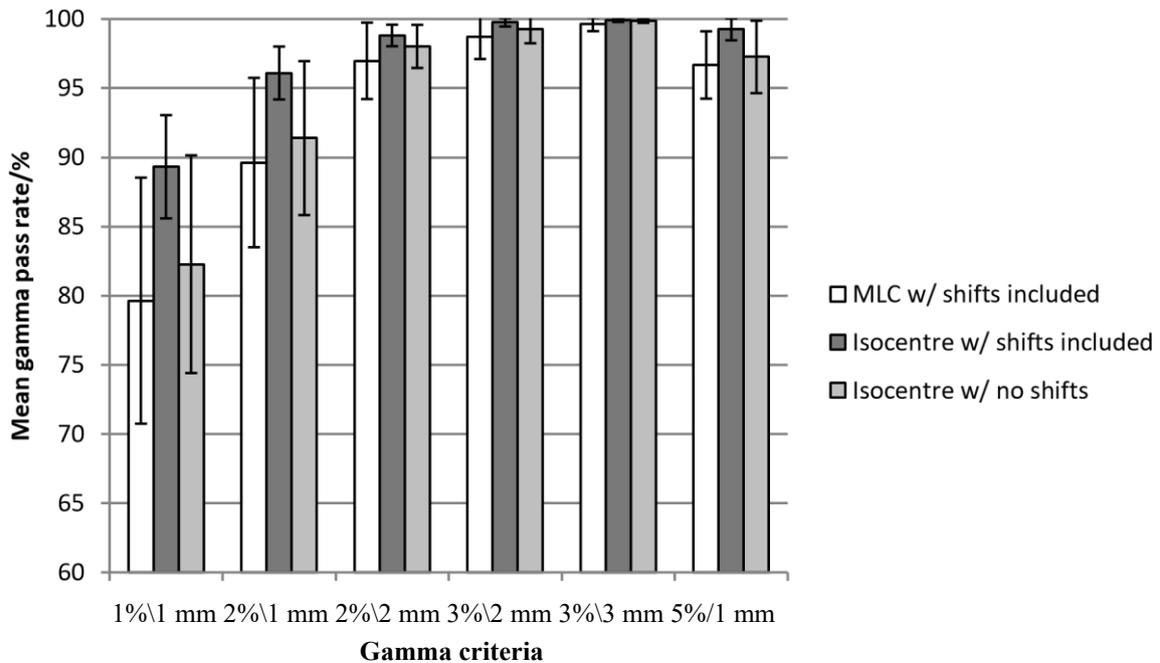

**Fig. 7** *Mean gamma pass rates for plans with introduced errors (excluding unedited plan data), with phantom in sagittal orientation. Error bars represent 1 SD from the mean*

## *ROC curves*

MapCHECK in MapPHAN is generally proficient at detecting introduced MLC errors, as demonstrated in Figures 8-9. Sensitivity and specificity for OAR doses are comparable to those in the literature for PTV doses measured with MapCHECK and ArcCHECK [17]. The shapes of the curves in both coronal and sagittal orientations are relatively similar for their respective gamma criteria, meaning that the difference in Youden indices for varying false positive rates was relatively low. However, there was some variation in AUC values for the gamma criteria of 3%/2mm and 3%/3mm, as seen in Table 1. It is likely that variations in these trends are statistical in nature, given the high degree of precision for this type of QA, and the relatively low tolerances used by these sets of criteria. Hence, it can be concluded that neither phantom orientation produces superior sensitivity or specificity for the detection of introduced MLC errors.

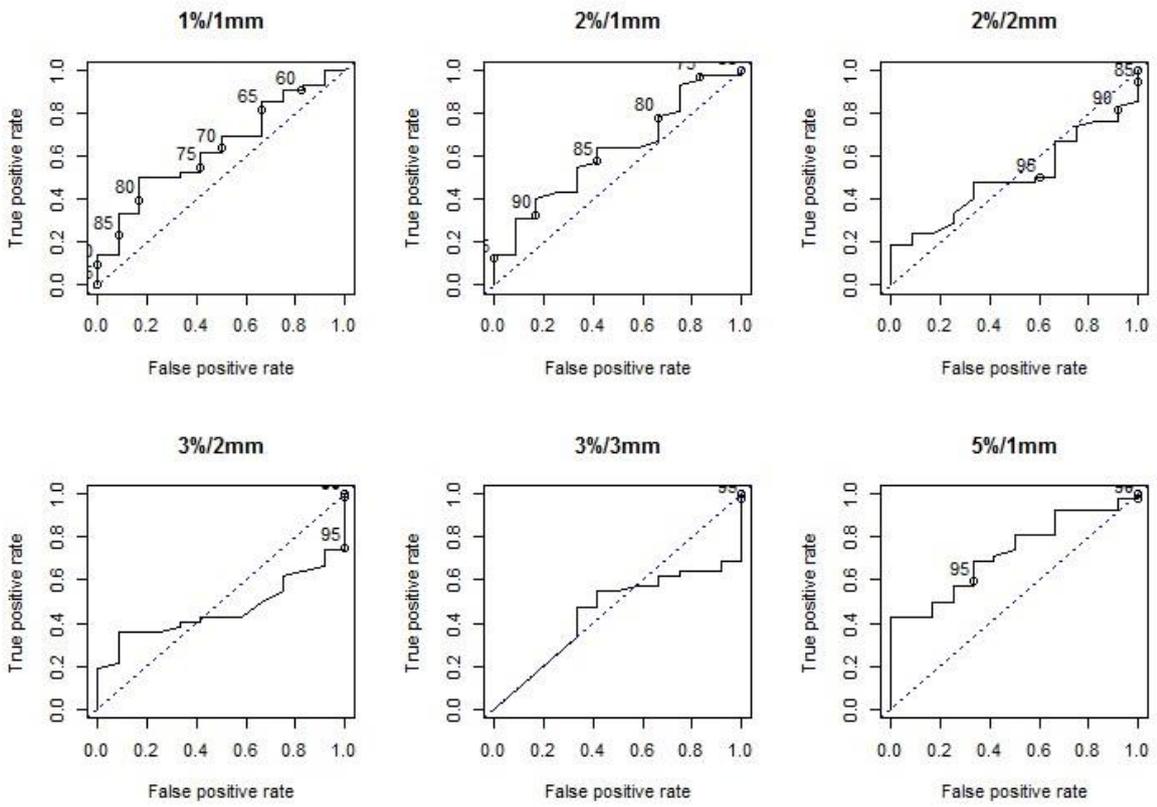

**Fig. 8** *ROC curves for introduced MLC shift errors with phantom in coronal orientation*

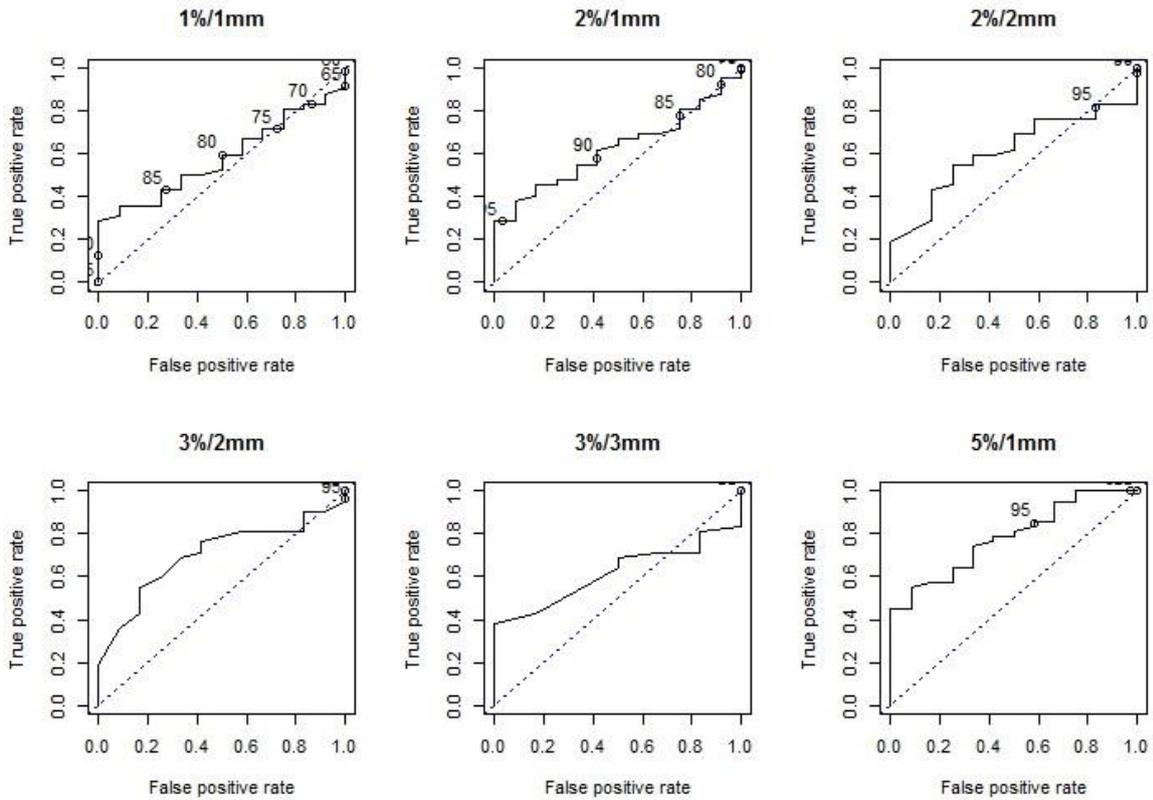

**Fig. 9** *ROC curves for introduced MLC shift errors with phantom in sagittal orientation*

| Error type | Phantom orientation | 1%/1mm | 2%/1mm | 2%/2mm | 3%/2mm | 3%/3mm | 5%/1mm |
|---|---|---|---|---|---|---|---|
| Isocentre | Coronal | 0.15 | 0.18 | 0.15 | 0.22 | 0.18 | 0.42 |
| Isocentre | Sagittal | 0.52 | 0.67 | 0.43 | 0.48 | 0.32 | 0.64 |
| MLC | Coronal | 0.64 | 0.62 | 0.50 | 0.46 | 0.45 | 0.73 |
| MLC | Sagittal | 0.58 | 0.62 | 0.61 | 0.70 | 0.62 | 0.78 |

**Table 1** *AUC for ROC curves for both isocentre and MLC shift errors, using all 6 gamma criteria*

The magnitude of isocentre shift errors used (1 mm) was based on current QA tolerances at Auckland Hospital (at the time of writing) but was too low to produce any statistically significant ROC curve data. This is evidence that the 1 mm QA tolerance is conservative, but in a favourable way, especially when considering the high geometrical precision required for SABR treatments and OAR sparing.

# Discussion

MLC leaf position errors have been studied on more complex IMRT plans, but to the authors' knowledge, this is the first study that looks at conformal arc plans for lung SABR specifically, with a focus on DVH metrics for OARs, rather than on the PTV. The lack of OAR analysis in the literature is an indication of the priority placed on PTV coverage during QA. For lung SABR, clinicians will often compromise PTV coverage to ensure OAR doses are within tolerance. This means that an ideal PSQA method should be sensitive to dosimetric discrepancies at the edge of the high dose region in the treatment plan.

The CT streak artefacts seen in Figure 1 have been reported previously by multiple authors [11]–[13]. Keeling *et al.* performed a similar analysis of the percent dose difference between measured and planned doses with no density overrides applied, reporting up to a 38% discrepancy between measured and calculated doses for 6 MV X-ray beams with 5 x 5 cm$^2$ fields that passed laterally through the diode array [12]. The method of optimising density overrides to reduce the angular dependency of the phantom to within $\pm$ 2% of measured values had equal effectiveness to other methods found in the literature, despite the lack of MV CT imaging capability [11]–[13].

The results in Figures 2-5 confirm that the maximum point dose received by an OAR is a function of the proximity of the OAR to the target volume. The most significant dosimetric increases observed were around the periphery of the PTV, meaning that OARs with tissue in these regions were most significantly affected by introduced errors. Normal tissue in these regions will experience a direct increase in primary beam output due to both introduced MLC and isocentre shift errors.

### Analysis of measurements

The results here suggest that 2%/1 mm with a tolerance between 80% and 90% is an appropriate set of criteria to use for QA, especially if one wishes to improve detection of OAR threshold failures. This has been shown by other authors; for instance, Kim *et al.* [1] found that 2%/1 mm, with passing rates of 80% and 90%, was the most suitable gamma criteria for PSQA of VMAT. However, it would be useful to perform a more detailed analysis of the 5%/1 mm criteria, in order to further confirm its suitability for QA. The detection accuracy of MapCHECK in MapPHAN to introduced errors was similar to that seen in the literature, based on the ROC analysis performed [1]–[3], [5], [22], [23], although in this study, the focus was on OARs rather than the PTV. For this reason, comparison with existing literature should be done tentatively, especially when considering the variation in devices, types of introduced errors, and gamma criteria between studies.

Overall, 5%/1 mm criteria with 95% tolerance appears to be uniquely suited to SABR QA. It aligns with the spatial accuracy required for SABR treatments, while being more lenient on dosimetric fluctuations, which are more prevalent in SABR plans due to their high dose gradients. However, current literature suggests there are few centres regularly using 5%/1 mm criteria, which means that there is little data available to validate its efficacy. This is unfortunate as the evidence here suggests that 5%/1 mm criteria may provide superior sensitivity and specificity to other criteria for SABR QA.

The ROC curves in this paper provide an indication of the best gamma criteria for PSQA of conformal arc plans. Generally, the performance of a single classifier can be quantified using the Youden index, which provides advantages over the AUC in that it is more specific to individual classifiers. Normally, two sets of gamma criteria are used in tandem to detect treatment errors. Based solely on a visual estimate of the Youden index for the ROC curves produced from the MLC results, an appropriate combination of criteria for clinical use could be a 90% tolerance with 2%/1 mm criteria, and a 95% tolerance with 5%/1 mm criteria. This could be used for QA with further optimization, depending on the needs of individual departments.

# Conclusion

MapCHECK in MapPHAN is a potential solution for replacing film as a more time-efficient method of SABR QA. Once angular dependency corrections are incorporated into the phantom model, it provides the ability to measure absolute dose, with ample detector resolution for measuring dose in high gradient regions. Based on the ROC analysis performed here, the phantom provides sufficient sensitivity and specificity for identifying introduced errors, over a wide range of gamma analysis parameters. The authors of this paper would recommend testing the use of 5%/1 mm gamma criteria with a 95% gamma pass rate tolerance for QA of lung SABR plans, in conjunction with 2%/1 mm, although these criteria should be optimised for individual clinics.

# Bibliography

[1]     J. Kim, S.-Y. Park, H. J. Kim, J. H. Kim, S.-J. Ye, and J. M. Park, "The sensitivity of gamma-index method to the positioning errors of high-definition MLC in patient-specific VMAT QA for SBRT," *Radiat. Oncol.*, vol. 9, no. 1, p. 167, 2014.

[2]     L. Vieillevigne, J. Molinier, T. Brun, and R. Ferrand, "Gamma index comparison of three VMAT QA systems and evaluation of their sensitivity to delivery errors," *Phys. Medica*, vol. 31, no. 7, pp. 720–725, 2015.

[3]     G. Yan, C. Liu, T. A. Simon, L.-C. Peng, C. Fox, and J. G. Li, "On the sensitivity of patient-specific IMRT QA to MLC positioning errors," *J. Appl. Clin. Med. Phys.*, vol. 10, no. 1, pp. 120–128, 2009.

[4]     A. Fredh, J. B. Scherman, L. S. Fog, and P. af Rosenschöld, "Patient QA systems for rotational radiation therapy: a comparative experimental study with intentional errors," *Med. Phys.*, vol. 40, no. 3, p. 31716, 2013.

[5]     B. Liang, B. Liu, F. Zhou, F. Yin, and Q. Wu, "Comparisons of volumetric modulated arc therapy (VMAT) quality assurance (QA) systems: sensitivity analysis to machine errors," *Radiat. Oncol.*, vol. 11, no. 1, p. 146, 2016.

[6]     C. P. Karger, D. Schulz-Ertner, B. H. Didinger, J. Debus, and O. Jäkel, "Influence of setup errors on spinal cord dose and treatment plan quality for cervical spine tumours: a phantom study for photon IMRT and heavy charged particle radiotherapy," *Phys. Med. Biol.*, vol. 48, no. 19, p. 3171, 2003.

[7]     A. K. Templeton, J. C. H. Chu, and J. V Turian, "The sensitivity of ArcCHECK-based gamma analysis to manufactured errors in helical tomotherapy radiation delivery," *J. Appl. Clin. Med. Phys.*, vol. 16, no. 1, pp. 32–39, 2015.

[8]     T. Alharthi, E. M. Pogson, S. Arumugam, L. Holloway, and D. Thwaites, "Pre-treatment verification of lung SBRT VMAT plans with delivery errors: toward a better understanding of the gamma index analysis," *Phys. Medica*, vol. 49, pp. 119–128, 2018.

[9]     M. R. Folkert and R. D. Timmerman, "Stereotactic ablative body radiosurgery (SABR) or Stereotactic body radiation therapy (SBRT)," *Adv. Drug Deliv. Rev.*, vol. 109, pp. 3–14, 2017.

[10] T. Aland, T. Kairn, and J. Kenny, "Evaluation of a Gafchromic EBT2 film dosimetry system for radiotherapy quality assurance," *Australas. Phys. Eng. Sci. Med.*, vol. 34, no. 2, pp. 251–260, 2011.

[11] P. A. Jursinic, R. Sharma, and J. Reuter, "MapCHECK used for rotational IMRT measurements:
step-and-shoot, TomoTherapy, RapidArc," *Med. Phys.*, vol. 37, no. 6Part1, pp. 2837–2846, 2010.

[12] V. P. Keeling, S. Ahmad, and H. Jin, "A comprehensive comparison study of three different planar IMRT QA techniques using MapCHECK 2," *J. Appl. Clin. Med. Phys.*, vol. 14, no. 6, pp. 222–233, 2013.

[13] J. Zhang, "SU-GG-T-247: A New Method to Compensate Angular Dependency of MapCheck Device in Intensity Modulated Arc Therapy," *Med. Phys.*, vol. 37, no. 6Part19, p. 3242, 2010.

[14] G. Van Rossum and F. L. Drake Jr, *Python reference manual*. Centrum voor Wiskunde en Informatica Amsterdam, 1995.

[15] T. Sing, O. Sander, N. Beerenwinkel, and T. Lengauer, "ROCR: visualizing classifier performance in R," *Bioinformatics*, vol. 21, no. 20, pp. 3940–3941, 2005.

[16] R Core Team, "R: A Language and Environment for Statistical Computing." Vienna, Austria, 2016.

[17] E. M. McKenzie, P. A. Balter, F. C. Stingo, J. Jones, D. S. Followill, and S. F. Kry, "Toward optimizing patient-specific IMRT QA techniques in the accurate detection of dosimetrically acceptable and unacceptable patient plans," *Med. Phys.*, vol. 41, no. 12, p. 121702, 2014.

[18] M. Goitein, "Calculation of the uncertainty in the dose delivered during radiation therapy," *Med. Phys.*, vol. 12, no. 5, pp. 608–612, 1985.

[19] F. Albertini, E. B. Hug, and A. J. Lomax, "Is it necessary to plan with safety margins for actively scanned proton therapy?," *Phys. Med. Biol.*, vol. 56, no. 14, p. 4399, 2011.

[20] T. Waschek, S. Levegrun, W. Schlegel, M. Van Kampen, and R. Engenhart-Cabillic, "Target volume definition for three-dimensional radiotherapy of cancer patients with a fuzzy rule based system," in *Proceedings of IEEE 5th International Fuzzy Systems*, 1996, vol. 3, pp. 1718–1725.

[21] N. Shusharina, D. Craft, Y.-L. Chen, H. Shih, and T. Bortfeld, "The clinical target distribution: a probabilistic alternative to the clinical target volume," *Phys. Med. Biol.*, vol. 63, no. 15, p. 155001, 2018.

[22] M. Carlone, C. Cruje, A. Rangel, R. McCabe, M. Nielsen, and M. MacPherson, "ROC analysis in patient specific quality assurance," *Med. Phys.*, vol. 40, no. 4, 2013.

[23] W. Liu, X. Zhang, Y. Li, and R. Mohan, "Robust optimization of intensity modulated proton therapy," *Med. Phys.*, vol. 39, no. 2, pp. 1079–1091, 2012.

[24] K. Nithiyanantham, G. K. Mani, V. Subramani, L. Mueller, K. K. Palaniappan, and T. Kataria, "Analysis of direct clinical consequences of MLC positional errors in volumetric-modulated arc therapy using 3D dosimetry system," *J. Appl. Clin. Med. Phys.*, vol. 16, no. 5, pp. 296–305, 2015.

[25]    A. Agnew, C. E. Agnew, M. W. D. Grattan, A. R. Hounsell, and C. K. McGarry, "Monitoring daily MLC positional errors using trajectory log files and EPID measurements for IMRT and VMAT deliveries," *Phys. Med. Biol.*, vol. 59, no. 9, p. N49, 2014.

[26]    T. Depuydt, A. Van Esch, and D. P. Huyskens, "A quantitative evaluation of IMRT dose distributions: refinement and clinical assessment of the gamma evaluation," *Radiother. Oncol.*, vol. 62, no. 3, pp. 309–319, 2002.